\documentclass[11pt]{article}

\usepackage[utf8]{inputenc}
\usepackage{authblk}
\usepackage{hyperref}
\usepackage{xcolor}
\usepackage{tabularx}
\usepackage{pifont}    
\usepackage{graphicx}
\usepackage{amssymb}
\usepackage{amsmath}
\usepackage{capt-of}
\usepackage{xurl}

\title{Cell Behavior Video Classification Challenge,\\a benchmark for computer vision methods\\in time-lapse microscopy}

\author[1,2,*]{Raffaella Fiamma Cabini}
\author[3]{Deborah Barkauskas}
\author[4]{Guangyu Chen}
\author[4]{Zhi-Qi Cheng}
\author[5]{David E Cicchetti}
\author[6]{Judith Drazba}
\author[7]{Rodrigo Fernandez-Gonzalez}
\author[7]{Raymond Hawkins}
\author[6]{Yujia Hu}
\author[8]{Jyoti Kini}
\author[4]{Charles LeWarne}
\author[9,10]{Xufeng Lin}
\author[11]{Sai Preethi Nakkina}
\author[6]{John W Peterson}
\author[12]{Ayushi Singh}
\author[5]{Koert Schreurs}
\author[5]{Kumaran Bala Kandan Viswanathan}
\author[5]{Inge MN Wortel}
\author[4]{Sanjian Zhang}
\author[1,2]{Rolf Krause}
\author[13,$\dagger$]{Santiago Fernandez Gonzalez}
\author[1,2,14,*,$\dagger$]{Diego Ulisse Pizzagalli}

\affil[1]{Euler Institute, Faculty of Informatics, Università della Svizzera italiana, Via la Santa 1, Lugano, 6962, Switzerland}
\affil[2]{International Center for Advanced Computing in Medicine (ICAM), University of Pavia, Via Bassi 6, Pavia, 27100, Italy}
\affil[3]{Imaging Platform, ACRF INCITe Centre, Garvan Institute of Medical Research, 384 Victoria Street, Sydney, 2010, NSW, Australia}
\affil[4]{Tacoma School of Engineering \& Technology, University of Washington, 1900 Commerce Street, Tacoma, 98402-3100, WA, USA}
\affil[5]{Data Science, Institute for Computing and Information Sciences, Radboud University, Houtlaan 4, Nijmegen, 6525EC, Netherlands}
\affil[6]{Imaging Core, Lerner Research Institute, Cleveland Clinic, 9500 Euclid Avenue, Cleveland, 44195, Ohio, USA}
\affil[7]{Institute of Biomedical Engineering, University of Toronto, 170 College Street, Toronto, M5S 3G9, Ontario, Canada}
\affil[8]{Center for Research in Computer Vision, University of Central Florida, 4328 Scorpius Street, Orlando, 32816-2365, Florida, USA}
\affil[9]{Computational Biology Group, Data Science Platform, Garvan Institute of Medical Research, 384 Victoria Street, Sydney, 2010, NSW, Australia}
\affil[10]{School of Clinical Medicine, Faculty of Medicine and Health, University of New South Wales, High Street, Sydney, 2052, NSW, Australia}
\affil[11]{Department of Pathology and Laboratory Medicine, Perelman School of Medicine, University of Pennsylvania, 3400 Spruce Street, Philadelphia, 19104-4238, PA, USA}
\affil[12]{Department of Ophthalmology and Visual Sciences, SUNY Upstate Medical University, 750 E Adams Street, Syracuse, 13210, NY, USA}
\affil[13]{Institute for Research in Biomedicine, Faculty of Biomedical Sciences, Università della Svizzera italiana, Via Francesco Chiesa 5, Bellinzona, 6500, Switzerland}
\affil[14]{Theodore Kocher Institute, Faculty of Medicine, University of Bern, Freiestrasse 1, Bern, 3001, Switzerland}
\affil[*]{Corresponding authors. E-mails: \texttt{raffaellacabini@gmail.com} and \texttt{pizzad@usi.ch}}
\affil[$\dagger$]{Equal contribution}

\date{}

\begin{document}

\maketitle

\begin{abstract}
The classification of microscopy videos capturing complex cellular behaviors is crucial for understanding and quantifying the dynamics of biological processes over time. However, it remains a frontier in computer vision, requiring approaches that effectively model the shape and motion of objects without rigid boundaries, extract hierarchical spatiotemporal features from entire image sequences rather than static frames, and account for multiple objects within the field of view.

To this end, we organized the Cell Behavior Video Classification Challenge (CBVCC), benchmarking 35 methods based on three approaches: classification of tracking-derived features, end-to-end deep learning architectures to directly learn spatiotemporal features from the entire video sequence without explicit cell tracking, or ensembling tracking-derived with image-derived features.

We discuss the results achieved by the participants and compare the potential and limitations of each approach, serving as a basis to foster the development of computer vision methods for studying cellular dynamics.
\end{abstract}

\section*{Introduction}
Studying cellular behavior over time is crucial for uncovering dynamic processes underlying tissue development, remodeling, and disease progression. Live-cell imaging enables the acquisition of time-lapse videos that capture the behavior of cells in real time, including morphological changes, migration, and interaction with other cells or extracellular structures. Among these techniques, intravital microscopy (IVM) is particularly well-suited to capture cellular dynamics in their native microenvironments with high spatial and temporal resolution~\cite{sumen2004intravital, beltman2009analysing, pizzagalli2019characterization}.

A key step in quantifying time-lapse microscopy data is detecting where in space and when in time cells exhibit specific behavioral patterns. This process is crucial for understanding both single-cell mechanisms, such as internal programs and pathways that induce specific migration patterns~\cite{pizzagalli2022vivo, schienstock2022moving}, as well as systemic processes driving the spatial arrangement of multiple cells across tissues and organs~\cite{waite2011dynamic, singer1999cutaneous, hanna2015patrolling}.
Despite a growing number of bioimage analysis methods for cell segmentation and tracking, detecting where and when specific cell behaviors occur still largely relies on the manual inspection of microscopy videos. This approach is time-consuming, subjective, and prone to errors.

Recent advances in computer vision have introduced action recognition, a methodology encompassing a variety of techniques to automatically analyze the behavior of humans and detect different actions~\cite{reddy2013recognizing, herath2017going}. While there has been recent interest in adapting these methods for the study of cellular behavior~\cite{anandakumaran2022rapid, delgado2024automatic, lee2020deep, liu2024deep, pulfer2024transformer, pizzagalli2025quantifying, molina2022acme}, their translation to the bioimaging field is limited by the absence of public, annotated, and standardized datasets. Furthermore, the classification of cell behavior from microscopy data presents unique challenges compared to human behavior in camera-acquired videos. These challenges include the lack of rigid shapes, large deformations from one frame to the next, and the absence of fixed key points (e.g., eyes, hands, nose). Moreover, cell behavior classification typically requires analyzing the entire video sequence rather than individual frames. In IVM, these aspects are further complicated by the complex tissue microenvironment present in the field of view, which contains multiple cells and extracellular structures that reduce the performance of automated cell segmentation and tracking~\cite{beltman2009analysing, pizzagalli2022cancol, schienstock2025cecelia}. Together, these factors necessitate advanced computational methods capable of deriving and processing spatiotemporal features from highly heterogeneous time-lapse microscopy videos.

The availability of public datasets and the organization of scientific challenges have been instrumental in developing, translating, and applying computer vision techniques for object detection, segmentation, and tracking in bioimaging~\cite{mavska2023cell, ulman_objective_2017, ma2024multimodality}. Within this context, we present the Cell Behavior Video Classification Challenge (CBVCC), explicitly aimed at benchmarking and improving computer vision methods for classifying time-lapse microscopy videos that capture diverse cell behaviors.

Specifically, we evaluated the ability of methods to: i) identify videos where cells exhibit sudden changes in migration direction, ii) distinguish these from videos showing cell with consistent, linear movement, stationary cells, and videos containing only background. Detecting changes in movement direction is a fundamental pattern, that is relevant from both a biological perspective (as it might indicate acute tissue damage, the formation of a strong chemotactic gradient, or migration schemes like scanning yielding extensive tissue monitoring), and from a computational perspective (as any method capable of classifying more complex behaviors should first be able first to identify this basic pattern). 

By offering a curated dataset and a well-defined classification framework, this challenge contributes to bridging the gap between biological imaging and computer vision, with the final aim of improving tools to better understand the intricate dynamics of living systems.

\section*{Results}
\subsection*{Design of the CBVCC challenge}
The CBVCC Challenge, organized between September 2024 and January 2025, consisted of two phases: an initial validation phase for method development and optimization (using a training and validation dataset), and a test phase for standardized performance evaluation on a separated test dataset. Participants had access to training and validation data for 28 days and the test data for 7 days.

Participants submitted their results for both the validation and test phases through the challenge website: \url{https://www.dp-lab.info/cbvcc/}. During the validation phase, teams could make multiple submissions, allowing them to refine their methods based on real-time leaderboard feedback. In contrast, the test phase was strictly reserved for performance evaluation and final ranking, allowing only one submission per team. Neither the validation nor the test datasets were intended for training.

Participants were required to adhere to strict data usage policies: external datasets were allowed only if unrelated to the specific classification task defined for the CBVCC challenge. Only automated or semi-automated methods were allowed. Manual annotation of the test datasets was strictly prohibited.

In total, 35 methods were evaluated through this challenge. 7 teams, from 7 different institutions, successfully submitted their method in both the validation and test phases (GIMR, LRI Imaging Core, QuantMorph, USI, the University of Washington, Radboud University, and the University of Central Florida). An additional 28 methods were submitted after the official challenge ended. These employed of-the-shelf deep learning (DL) architectures, and the analysis of their results were included in a separate analysis.

\subsection*{Dataset}
To provide a sufficiently extended dataset to train and validate the submitted methods, we curated and labeled 300 2D video-patches. This dataset was derived from 48 independent 3D IVM acquisitions in a murine flank skin model and captured the migration of T lymphocytes under inflammatory conditions~\cite{pizzagalli2025systematic, norman2021dynamic} (Fig.~\ref{fig:fig1}A-C).

Each acquisition lasted 30 minutes, with frames captured every 60 seconds (resulting in 31 frames per video), a field of view of 263.1 × 264.8 $\mu m^2$ $\pm$ 8.2 × 10.2 $\mu m^2$ (mean $\pm$ standard deviation) and a thickness of 138.9 $\mu m$ $\pm$ 26.6 $\mu m$ (slice thickness: 2.0 $\mu m$ $\pm$ 0.2 $\mu m$).

From these videos, we extracted 300 video-patches of reduced size. Each video-patch was labeled based on the behavior of the cell located at its spatial center at the temporal midpoint. The labelling design supports future integration with sliding-window strategies, which could enable dense scanning of the original full-size IVM videos and the generation of spatially and temporally resolved maps of cellular actions across the full field of view. Specifically, 120 video-patches were centered on cells that exhibited sudden changes in their direction of movement (Class 1), and 180 video-patches were centered on cells without sudden changes direction (moving linearly, remaining stationary) or background areas without visible cells (Class 0) (Fig.~\ref{fig:fig1}D). Each video-patch was a 2D projection along the z-axis of the 3D acquisition, with a field of view of 40.0 × 40.0 $\mu m^2$ (matrix size: 50 × 50 pixels), lasting 19 minutes, with frames captured every 60 seconds (resulting in 20 frames per video-patch). All video-patches were adjusted to a common contrast range and preprocessed to ensure a uniform pixel size of 0.8 $\mu m$.

Labeled video patches were partitioned and provided to participants of the CBVCC as follows: 210 (70\%) for training set, 30 (10\%) for validation set, and 60 (20\%) for test set (Fig~\ref{fig:fig1}E). Each subset included video-patches extracted from different and independent IVM videos. The labels of the test-set were not provided to participants until the challenge concluded. The number of cells and signal-to-noise ratio (SNR) were different for each video-patch as outlined in Fig.~\ref{fig:fig1}F.

{\centering\includegraphics[width=0.885\textwidth]{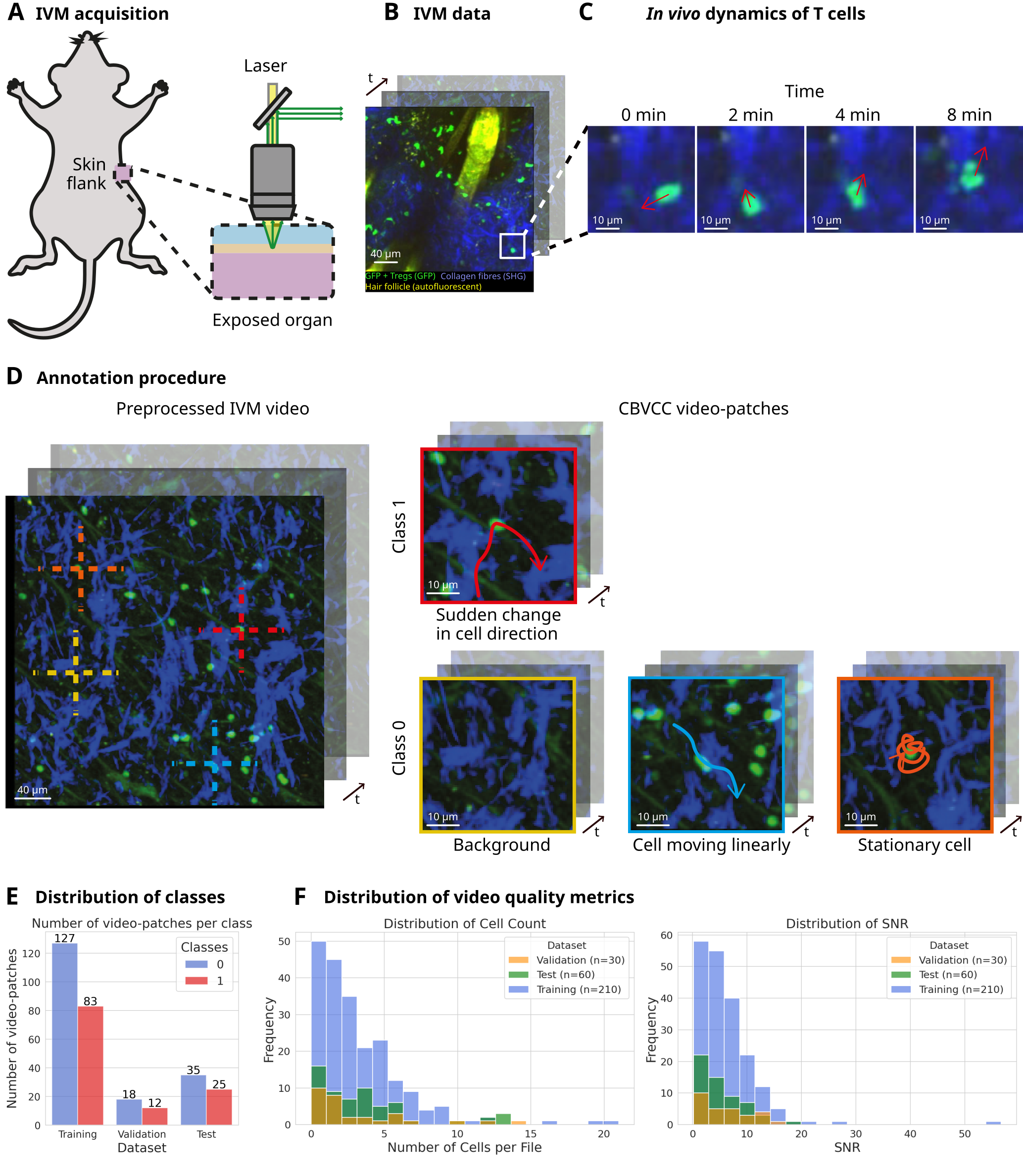}
\captionof{figure}{ \footnotesize \textbf{IVM data acquisition, annotation, and dataset composition.}\\ \textbf{A.} Simplified representation of the IVM video acquisition. The flank skin of anesthetized and immobilized mice is surgically exposed and stabilized using a specialized imaging window. Imaging is performed using a multiphoton microscope with point-wise excitation. \textbf{B.} Example of acquired IVM video, consisting of a z-stack of parallel image planes over time across multiple fluorescence channels. \textbf{C.} Cropped sequence of frames highlighting the dynamic movement of a migrating T cell. \textbf{D.} Schematic representation of the annotation process: an operator semi-manually navigated through the video frames to select coordinates for video-patch extraction. Class 1 video-patches were annotated to include instances where the central cell in the middle frame changes its direction of movement. Class 0 video-patches were annotated to include cells that move in a linear trajectory, stationary cells, or some background regions with no visible cells. \textbf{E.} Distribution of the number of video-patches annotated as Class 0 and Class 1, divided among the three datasets. \textbf{F.} Distribution of the number of cells and the signal-to-noise ratio (SNR) per video-patch across the three datasets.}\label{fig:fig1}}

\subsection*{Summary of participating algorithms}
The methods that participated in the CBVCC challenge can be categorized into two main approaches: tracking-based methods and fully end-to-end DL methods (Fig.~\ref{fig:fig2}A-B). Tracking-based methods generally analyze the trajectory of a single cell, whereas end-to-end DL methods process the entire video at once.  A summary of the participating algorithms is provided in Table~\ref{tab:tab1}.

Tracking-based approaches (Fig.~\ref{fig:fig2}A) first segmented and tracked cells, then classified them based on motility features. Various DL-based segmentation methods were used, including an off-the-shelf or retrained CellPose model~\cite{stringer2025cellpose3} and a Detectron2-based detection module~\cite{wu2019detectron2}. Tracking was performed using different techniques: some teams used TrackPy~\cite{allan2021soft} to link cell centroid coordinates across frames based on distance thresholds, while another linked cell centroids with the minimum Euclidean distance. One team utilized the Segment Anything Model 2 (SAM 2)~\cite{ravi2024sam} to simultaneously perform segmentation and tracking, enabling trajectory generation without the need for manually labeled ground truth, in contrast to the other segmentation methods that depended on labeled data for training. Following segmentation and tracking, classification involved extracting motility features from cell trajectories, either using pre-defined track descriptors or features learned by a DL model. These features were then fed into behavior classifiers, including traditional machine learning models (logistic regression) and DL architectures (CNNs, LSTMs, and attention-based networks) with multi-layer perceptron for final classification.

End-to-end DL methods (Fig.~\ref{fig:fig2}B) directly classify videos using neural networks, bypassing explicit segmentation or tracking steps. The methods that participated in the challenge primarily leveraged convolutional neural networks (CNNs) or attention mechanisms. CNN-based models employed 3D convolutional layers to simultaneously process the spatial and temporal information of videos~\cite{ji20123d}, focusing on local feature extraction. Attention-based models, such as the Video Shifted Window Transformer (Swin)~\cite{liu2021swin}, used multi-head self-attention to capture global dependencies across video sequences, making them more effective for long-term behaviors but computationally more expensive. Both these methods are particularly promising because they automatically process the entire video, which is beneficial in scenarios where tracking is difficult or prone to errors.

The LRI Imaging Core team combined these approaches by using cell trajectories to crop the video around regions of interest, followed by classification. Classification was performed using a combination of two DL models: the Animation Analyzer, which processed the cropped video data, and the Pattern Recognizer, which examined 2D cell profile projections in individual frames~\cite{hu2023labgym, goss2024quantifying}.

All the participating tracking-based approaches selected a trajectory of interest based on predefined rules before classification, except for the winning method of the CBVCC challenge, which deviated from this practice by using a DL attention mechanism to automatically select the most informative trajectories within each video. Except for the Computational Immunology team, all tracking-based approaches required additional annotations beyond those provided by the CBVCC challenge. These included manually labeled ground-truth segmentations, single-cell selection, or annotation of all trajectories within the field of view. In contrast, end-to-end DL models did not require such extra annotations. However, these models were prone to overfitting, which was only partially mitigated through extensive data augmentation and model parameter reduction. Tracking-based methods, in comparison, were significantly less affected by overfitting. 

The extended description of each participating algorithm is provided in Supplementary Material 2 - Description of the participating algorithms.

\begin{table}[h]
\centering
\resizebox{\textwidth}{!}{%
\begin{tabular}{c l p{2.5cm} p{1.6cm} p{2.2cm} l p{2.0cm} p{2.4cm} p{1.6cm} l p{3.2cm}}
\hline
\textbf{Rank} & \textbf{Score} & \textbf{Team name} & \textbf{End-to-end DL} & \textbf{Tracking} & \textbf{Segmentation} & \textbf{Cell selection} & \textbf{Extra-annotations} & \textbf{Pre-trained} & \textbf{Ensemble} & \textbf{Architecture} \\ \hline
1 & 0.922 & GIMR & \ding{55} & \ding{51} & \ding{51} & \ding{55} & \ding{51} (manual seg.) & \ding{55} & \ding{51} & 1DCNN + Attention module\\
2 & 0.853 & LRI Imaging Core & \ding{55} & \ding{51} & \ding{51} & \ding{51} & \ding{51} (manual seg.) & \ding{55} & \ding{55} & 2DCNN + LSTM \\
3 & 0.835 & QuantMorph & \ding{55} & \ding{51} & \ding{51} & \ding{51} & \ding{51} (removing extra cells) & \ding{51} & \ding{55} & 1DCNN + LSTM\\
4 & 0.815 & dp-lab & \ding{51} & \ding{55} & \ding{55} & \ding{55} & \ding{55} & \ding{55} & \ding{55} & 3DCNN\\
5 & 0.752 & UWT-SET & \ding{51} & \ding{55} (for data augm.) & \ding{55} & \ding{55} & \ding{55} & \ding{55} & \ding{55} & Swin-Tiny\\
6 & 0.749 & Computational Immunology & \ding{51} & \ding{51} & \ding{51} & \ding{51} & \ding{55} & \ding{51} & \ding{51} & Ensemble of logistic regression and 3DCNN\\
7 & 0.716 & BioVision & \ding{51} & \ding{55} & \ding{55} & \ding{55} & \ding{55} & \ding{51} & \ding{55} & Swin\\
\hline
\end{tabular}%
}
\caption{\footnotesize Summary of the characteristics of the participating methods in the final test evaluation. We report whether the training required extra annotations beyond those provided in the challenge, if participants used pretrained models, if an ensemble approach was applied, if tracking and segmentation were used, if the authors selected a single cell or processed all cells in the field of view, and the type of model used for the final classification.}
\label{tab:tab1}
\end{table}

{\centering\includegraphics[width=\textwidth]{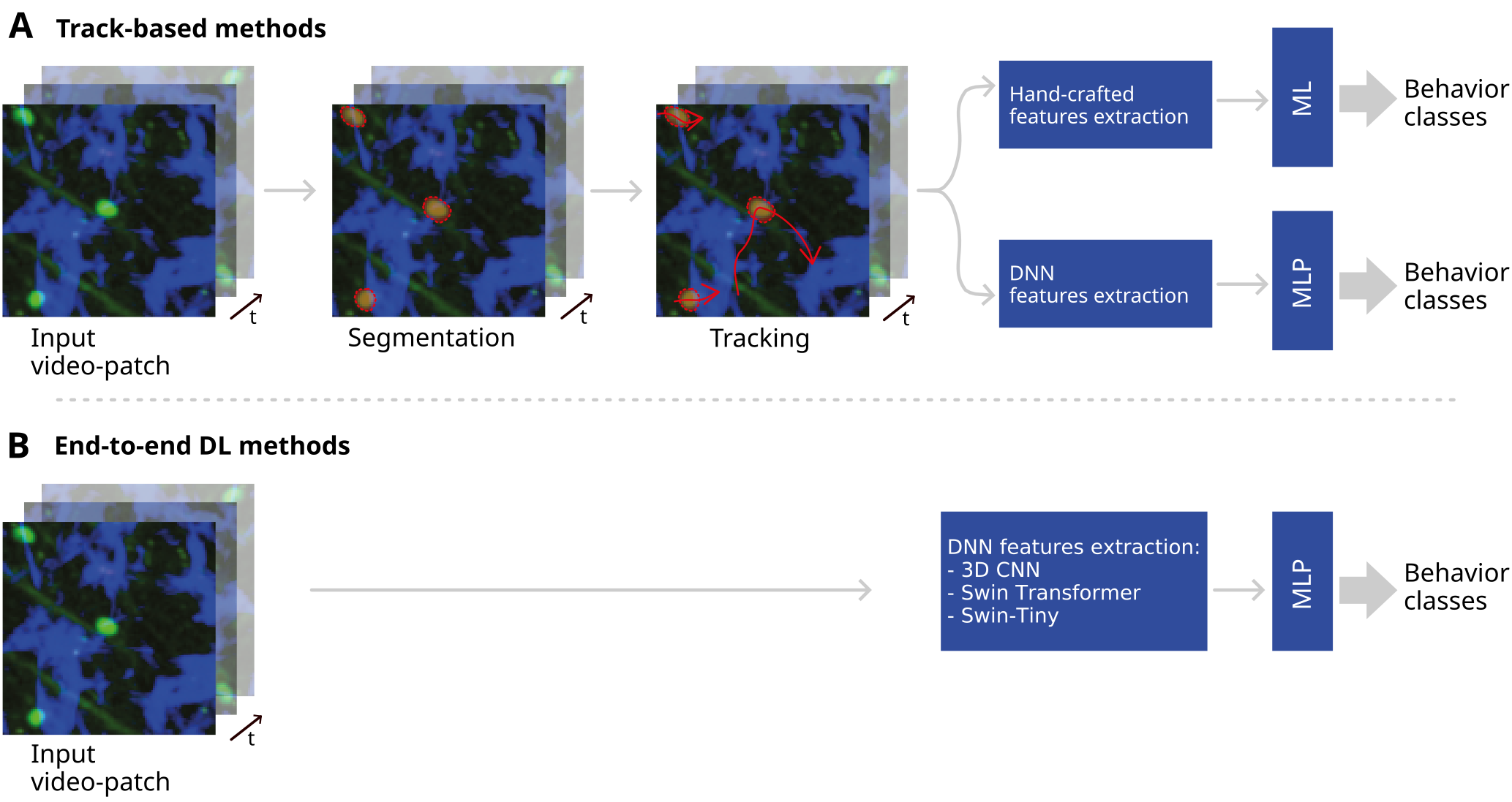}
\captionof{figure}{\footnotesize \textbf{Approaches for the classification of CBVCC video-patches.} \textbf{A.} Track-based methods involve segmentation, tracking, feature extraction using either handcrafted features or deep neural networks (DNNs), and classification through machine learning (ML) models or a multilayer perceptron (MLP). \textbf{B.} End-to-end DL methods include an automatic feature extraction module that processes video-patches using a DNN, followed by an MLP for class prediction.
}\label{fig:fig2}}

\subsection*{Performances of participating algorithms}
The scores of the participating methods in the video-patches classification task are divided between validation phase (Table~\ref{tab:tab2}A) accounting the best score from multiple submissions on the validation set, and the test phase (Table~\ref{tab:tab2}B) presenting the score from a single submission on the previously unseen and independent test set.

On the validation set, all methods achieved an overall score between 0.805 and 0.970. AUC values ranged from 0.819 to 0.977, precision varied between 0.750 and 0.923, recall ranged from 0.750 to 1.000, and balanced accuracy ranged from 0.792 to 0.972. The Computational Immunology team led in most of the metrics, with an AUC of 0.981, precision of 1.000, recall of 0.917, and balanced accuracy of 0.958, achieving an overall score of 0.968.

On the test set, performance varied, with overall scores ranging from 0.716 to 0.922. AUC values ranged from 0.784 to 0.944, precision varied between 0.588 and 0.840, recall ranged from 0.560 to 1.000, and balanced accuracy ranged from 0.700 to 0.914. Two main trends were observed: stable or slightly improved performances from validation to test evaluation (4 of 7 methods), or evident decline (3 of 7 methods).  Notably, the GIMR method ranked first, with an AUC of 0.944, precision of 0.806, recall of 1.000, and balanced accuracy of 0.914, achieving an overall score of 0.922. In contrast, the Computational Immunology method experienced a drop, with its AUC decreasing to 0.827, precision to 0.588, recall to 0.800, balanced accuracy to 0.700, and overall score to 0.749. 

The ROC curves for the methods on both the validation and test datasets (Fig.~\ref{fig:fig3}A-B) with the overall scores reported in (Table~\ref{tab:tab2}A-B), confirming the overall ranking and performance variations between the validation and test sets.

To better understand the dependence of the methods’ performance on video complexity and quality metrics, we further analyzed the impact of the number of cells present in the video-patches on the final score (Fig.~\ref{fig:fig3}C). The results indicate a general trend of decreasing performance as the number of cells increases, which can be attributed to the additional variability introduced by neighboring cells and a higher risk of tracking errors. However, the winning GIMR team and the fourth-ranked dp-lab team maintained relatively stable performance. We also evaluated the effect of SNR on the final score (Fig.~\ref{fig:fig3}D), which revealed greater variability in performance across teams. The top-ranked GIMR team and the second-ranked LRI team demonstrate the most consistent results for all SNR values.

\begin{table}
\centering
\resizebox{\textwidth}{!}{%
\begin{tabular}{c l c c c c c}
\hline
\textbf{Rank} & \textbf{Team name} & \textbf{AUC} & \textbf{Precision} & \textbf{Recall} & \textbf{Balanced Accuracy} & \textbf{Overall score} \\
\hline
\multicolumn{7}{c}{\textbf{A. Validation set}} \\
1 & Computational Immunology & \textbf{0.981} & \textbf{1.000} & \textbf{0.917} & \textbf{0.917} & \textbf{0.968} \\
2 & UWT-SET & 0.949 & 0.846 & \textbf{0.917} & 0.903 & 0.913 \\
3 & GIMR & 0.931 & 0.786 & \textbf{0.917} & 0.875 & 0.888 \\
4 & QuantMorph & 0.856 & 0.900 & 0.750 & 0.847 & 0.842 \\
5 & dp-lab & 0.889 & 0.750 & 0.750 & 0.792 & 0.814 \\
6 & BioVision & 0.775 & \textbf{1.000} & 0.667 & 0.833 & 0.810 \\
7 & LRI Imaging Core & 0.819 & 0.818 & 0.750 & 0.819 & 0.805 \\
\hline
\multicolumn{7}{c}{\textbf{B. Test set}} \\
1 & GIMR & \textbf{0.944} & 0.806 & \textbf{1.000} & \textbf{0.914} & \textbf{0.922} \\
2 & LRI Imaging Core & 0.861 & \textbf{0.840} & 0.840 & 0.863 & 0.853 \\
3 & QuantMorph & 0.887 & 0.710 & 0.880 & 0.811 & 0.835 \\
4 & dp-lab & 0.880 & 0.760 & 0.760 & 0.794 & 0.815 \\
5 & UWT-SET & 0.784 & 0.679 & 0.760 & 0.751 & 0.752 \\
6 & Computational Immunology & 0.827 & 0.588 & 0.800 & 0.700 & 0.749 \\
7 & BioVision & 0.787 & 0.737 & 0.560 & 0.709 & 0.716 \\
\hline
\end{tabular}%
}
\caption{\footnotesize Final scores of the participating methods in the video-patches classification task on the validation set \textbf{A} and test set \textbf{B}. The table reports performance metrics, including AUC, precision, recall, balanced accuracy, and the overall score. For the validation set, only each team’s best performing model, selected from their multiple submissions, was considered. The best results for each metric are highlighted in bold.}
\label{tab:tab2}
\end{table}

{\centering\includegraphics[width=0.80\textwidth]{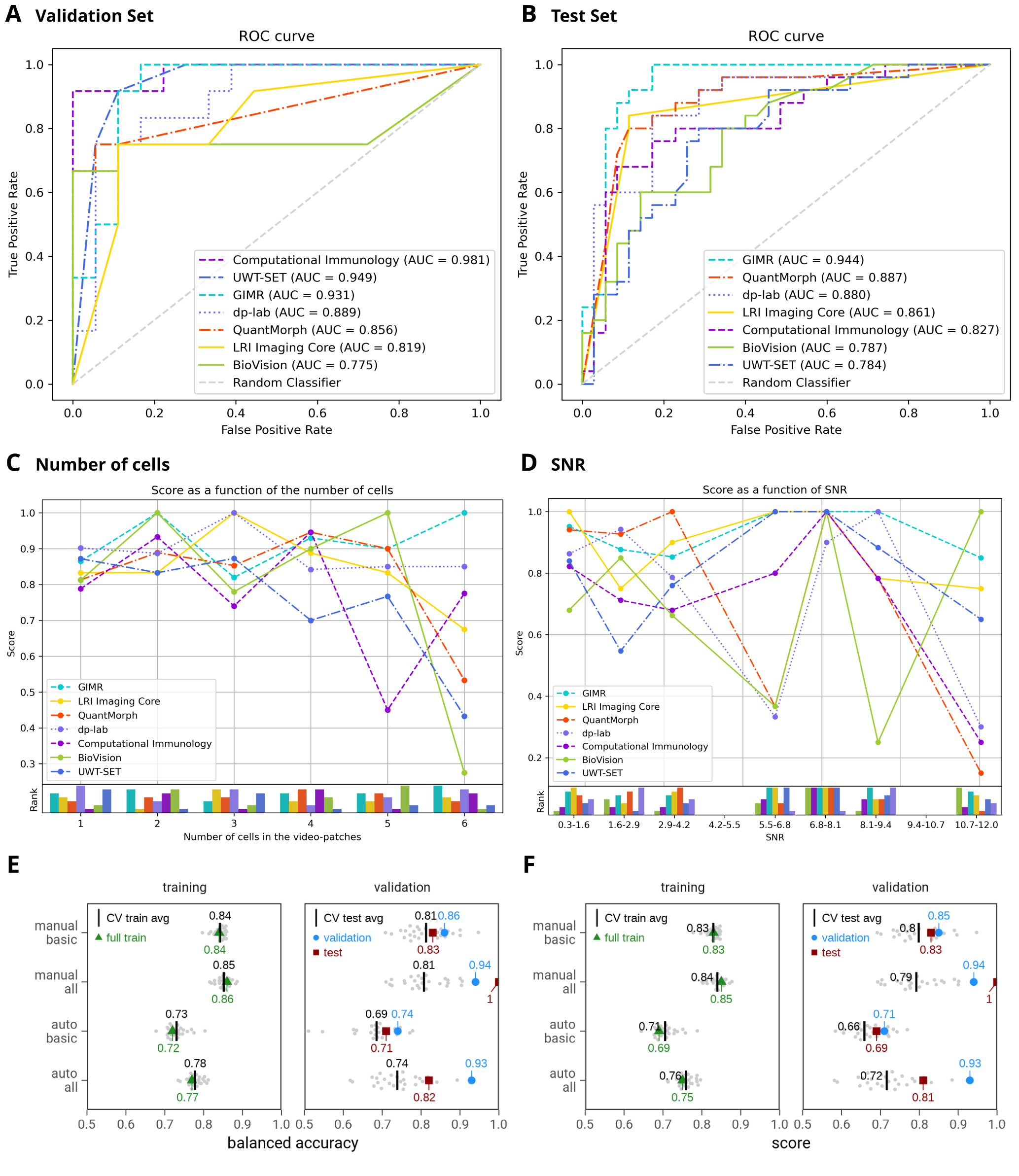}
\captionof{figure}{\footnotesize \textbf{Performance Evaluation.} \textbf{A.} ROC curve of the participating methods for the video-patch classification task on the validation set. The curves represent the performance of only the best model from each team, selected from their multiple submissions. \textbf{B.} ROC curve of the participating methods on the test set, illustrating the trade-off between sensitivity and specificity and highlighting performance differences with the validation set. \textbf{C.} Performance on the test set as a function of the number of cells in the video-patch. The bar plot below ranks each team based on the number of cells. \textbf{D.} Performance trend on the test set as a function of the signal-to-noise ratio (SNR) of the video-patch. A maximum SNR of 12.0 was used to maintain an adequate number of video-patches. The accompanying bar plot ranks each team based on the SNR. \textbf{E.} Baseline logistic regression performance in terms of balanced accuracy. \textbf{F.} Baseline logistic regression performance in terms of overall score. In E and F gray dots represent results from 5 repeats of 5-fold cross-validation (25 models), with average performance (\textbar) shown on the training (left) and validation (right) folds. Colored dots represent results from training on the entire training dataset and evaluation on the CBVCC training set ($\blacktriangle$), validation set ($\bullet$), and test set ($\blacksquare$). “basic” indicates the set of standard motility features, “all” the extended set including the two additional hand-crafted features, auto automated tracks, and manual manually annotated tracks.}\label{fig:fig3}}

\subsection*{Benchmarking state-of-the-art DL video classification models}
To complement the challenge results, we conducted an independent benchmark of state-of-the-art end-to-end DL models for video classification (Table~\ref{tab:tab3}). This evaluation was conducted on the test phase dataset of the CBVCC, using standard baseline architectures and training procedures without custom modifications nor model tuning. The goal of this evaluation was to establish a reference performance level and provide insights into how current DL leading approaches perform on this dataset under standardized conditions. The evaluation includes spatiotemporal CNN-based methods (X3D~\cite{feichtenhofer2020x3d}, I3D~\cite{carreira2017quo}, R2plus1D~\cite{tran2018closer}, CSN~\cite{tran2019video}, TSM~\cite{tsm2019}, SlowOnly and SlowFast~\cite{feichtenhofer2019slowfast}, TSN~\cite{wang2016temporal}), transformer-based methods (Swin~\cite{wang2016temporal}, ViTClip~\cite{arnab2021vivit}) and a recurrent neural network (TRN~\cite{xu2019temporal}). For CNN-based models, we employed various backbone architectures, including ResNet-based architectures~\cite{he2016deep} with 32 (R32), 50 (R50), 101 (R101), and 152 layers (R152), as well as MobileNetV2~\cite{sandler2018mobilenetv2}.

Spatiotemporal CNN-based approaches performed competitively, with TSM-R50 achieving a score of 0.709, and SlowOnly-R101 reaching a score of 0.726. Transformer-based methods such as Swin-small and ViTClip-large demonstrated moderate results, with Swin-small achieving a score of 0.596, and ViTClip-large reaching a score of 0.626. The recurrent neural network TRN exhibited a lower score of 0.582.

\begin{table}
\centering
\resizebox{\textwidth}{!}{%
\begin{tabular}{c l c c c c c}
\hline
\textbf{Rank} & \textbf{DL Model} & \textbf{AUC} & \textbf{Precision} & \textbf{Recall} & \textbf{Balanced Accuracy} & \textbf{Score} \\
\hline
1 & SlowOnly-R101~\cite{feichtenhofer2019slowfast} & 0.750 & 0.535 & 0.920 & 0.674 & \textbf{0.726} \\
2 & TSM-R50~\cite{tsm2019} & \textbf{0.781} & 0.682 & 0.600 & \textbf{0.700} & 0.709 \\
3 & SlowFast-R50~\cite{feichtenhofer2019slowfast} & 0.744 & 0.625 & 0.600 & 0.671 & 0.677 \\
4 & X3D-S~\cite{feichtenhofer2020x3d} & 0.769 & \textbf{0.750} & 0.360 & 0.637 & 0.657 \\
5 & CSN~\cite{tran2019video} & 0.669 & 0.417 & \textbf{1.000} & 0.500 & 0.651 \\
6 & TSM-R152~\cite{tsm2019} & 0.693 & 0.600 & 0.600 & 0.657 & 0.648 \\
7 & TSM-R101~\cite{tsm2019} & 0.726 & 0.600 & 0.480 & 0.626 & 0.631 \\
8 & ViTClip-large~\cite{arnab2021vivit} & 0.665 & 0.667 & 0.480 & 0.654 & 0.626 \\
9 & ViTClip-base~\cite{arnab2021vivit} & 0.669 & 0.533 & 0.640 & 0.620 & 0.626 \\
10 & Swin-small~\cite{liu2021swin} & 0.631 & 0.517 & 0.600 & 0.600 & 0.596 \\
11 & Swin-tiny~\cite{liu2021swin} & 0.632 & 0.484 & 0.600 & 0.571 & 0.584 \\
12 & TRN~\cite{xu2019temporal} & 0.747 & \textbf{0.750} & 0.120 & 0.546 & 0.582 \\
13 & TSM-MobileNetV2~\cite{tsm2019} & 0.702 & 0.562 & 0.360 & 0.580 & 0.581 \\
14 & SlowOnly-R50~\cite{feichtenhofer2019slowfast} & 0.726 & 0.583 & 0.280 & 0.569 & 0.577 \\
15 & SlowFast-R101~\cite{feichtenhofer2019slowfast} & 0.655 & 0.556 & 0.400 & 0.586 & 0.570 \\
16 & Swin-base~\cite{liu2021swin} & 0.578 & 0.436 & 0.680 & 0.526 & 0.560 \\
17 & TSN-R50~\cite{wang2016temporal} & 0.603 & 0.500 & 0.480 & 0.569 & 0.551 \\
18 & R2plus1D-R34~\cite{tran2018closer} & 0.425 & 0.556 & 0.200 & 0.543 & 0.430 \\
19 & I3D~\cite{carreira2017quo} & 0.663 & 0.000 & 0.000 & 0.486 & 0.363 \\
\hline
\end{tabular}%
}
\caption{\footnotesize Benchmark results of state-of-the-art end-to-end DL models for video-patches classification on the test set. The table presents performance metrics such as AUC, precision, recall, balanced accuracy, and the overall score. The best-performing models for each metric are highlighted in bold.}
\label{tab:tab3}
\end{table}

\subsection*{Baseline Comparison with Logistic Regression}
To setup a baseline for comparison, we evaluated the performance of a simple track classifier on the CBVCC dataset, aiming to: i) assess the impact of the tracking step on the overall performance by comparing an off-the-shelf automated tracking approach with manually annotated tracks, and ii) examine how generalization performance varies depending on the specific composition of the dataset used for testing.

The baseline model consists of the following steps. First, images were segmented using Cellpose~\cite{stringer2025cellpose3}, followed by automated track linking with minimal parameter tuning. From the resulting tracks, we extracted a set of basic motility features and two additional hand-crafted features (described in Methods and Supplementary Material 1). The extracted features were then used as input for a logistic regression classifier. We evaluated the model’s generalization performance using both the CBVCC dataset partition (validation set N = 30, test set N = 60) and a 5-fold cross-validation, repeating the entire CV procedure 5 times.

On the original partitioning of the dataset (train, validation, and test sets) provided along the CBVCC challenge, the model including both basic and hand-crafted features had a balanced accuracy of 0.93 on the validation data set (Fig.~\ref{fig:fig3}E-F, “auto\_all”), which dropped to 0.74 without hand-crafted features (Fig.~\ref{fig:fig3}E-F, “auto\_basic”). Performance on the test dataset was lower (“auto\_all” 0.82, “auto\_basic” 0.71). Overall, training and test performances were similar, suggesting a limited tendency to overfit. In the cross-validation partitioning, models trained on different folds or the full training set had similar results (Fig.~\ref{fig:fig3}E left).  However, we observed a moderate change in validation accuracy (0.65 to 0.85, Fig.~\ref{fig:fig3}F, right), which reflected differences in the validation set, which was randomly chosen for each fold.

To investigate whether performance limitations were due to tracking errors rather than the simplicity of the logistic regression model, we further repeated the experiments using manually annotated tracks. With high-quality manual tracks, the classifier achieved significantly higher performances, achieving validation/test balanced accuracies of 0.94/1.00 for the model using manual tracks and the complete set of features (Fig.~\ref{fig:fig3}E, “manual\_all”), and 0.86/0.83 for the model using manual tracks and the basic set of features (Fig.~\ref{fig:fig3}A, “manual\_basic”). Exploratory analyses suggested that the extracted features could capture most of the variance between class 0 and class 1 samples (Fig. S1-2). Finally, we found that even a single trained model varied substantially in performance across multiple instances of the automated tracking pipeline on the same test set (Fig. S3-4). These results highlight the sensitivity of the track-based classifier to errors arising from automated tracking.

\section*{Discussion}
We presented the CBVCC challenge and analyzed the performance of participating algorithms. The challenge focused on distinguishing videos where cells exhibit sudden directional changes from those showing consistent linear migration, stationary behavior, or background noise. In the proposed dataset, each video-patch was constructed such that the cell undergoing the directional change was located at the spatial and temporal center of the clip. This labelling design was adopted with the aim of facilitating future integration with sliding-window approaches, which could enable dense scanning of the original full-size IVM videos and the generation of spatially and temporally resolved maps of cellular actions across the full field of view. Detecting changes in cell migration direction from microscopy videos is a computationally demanding task, requiring analysis over a sufficiently large temporal window rather than individual frames. Therefore, this task represents an optimal benchmark for evaluating methods that account for spatiotemporal features.

To support reproducible benchmarking, we provided a curated, publicly available dataset and established a standardized protocol for training, evaluation, and testing. Following current best practices~\cite{kavur2021chaos}, the challenge restricted participants to a single test set submission, preventing optimization on the test data and ensuring fair comparisons.

Participants mainly adopted two distinct approaches: track-based methods and end-to-end DL models. Track-based methods outperformed others on both the validation and test sets. This advantage probably arises from the lower number of parameters of the classifiers, which enables better generalization with smaller datasets. However, their performance decreased in situations where automated tracking was particularly challenging or prone to errors, highlighting their dependence on detection and tracking algorithms, which are active areas of research. Additionally, the classification of behaviors associated with changes in morphology or involving the interaction of multiple cells may require morphodynamical features, which are not easily captured by standard centroid tracking.

By contrast, end-to-end DL methods generally showed lower performance. Main challenges included overfitting, the limited size of the training set, and the inherent complexity of the proposed classification problem, which requires analyzing entire video sequences rather than relying on features from individual frames. Additionally, IVM poses further challenges due to background noise arising from the tissue microenvironment, and the presence of multiple moving objects within the field of view. Nevertheless, some teams achieved satisfactory generalization results by applying strategies such as video preprocessing, reducing the model complexity, and heavy data augmentation, mitigating overfitting despite the relatively small training set.

Alongside the challenge results, we performed an independent benchmark of state-of-the-art end-to-end DL models for video classification. In this benchmark, CNN-based architectures (e.g., SlowOnly-R101 and TSM-R50) achieved the highest performance. Transformer-based models (e.g., Swin-small and ViTClip-large) yielded moderate performance, while recurrent models (e.g., TRN) underperformed, suggesting that not all temporal architectures are equally suited to this task.

Only three out of seven teams participating in the challenge used transfer learning. Pretraining on large datasets (even from unrelated domains) followed by fine-tuning reduced training time and improved performance. One team pre-trained an end-to-end DL model on a human behavior dataset, another adapted the pretrained SAM2 model for automatic segmentation and tracking in microscopy images, initializing it with a blob detector as seed points generator. A third combined pretrained CellPose for segmentation with TrackPy for tracking. 

These results demonstrate that, despite differences across imaging domains, large datasets and pretrained models can be highly valuable when carefully adapted. This highlights the importance of initiatives like the CBVCC challenge, which provides high-quality, publicly available datasets. For these reasons, we hope the CBVCC dataset will serve not only as a benchmark but also as a resource for pretraining models that can be applied to a broader range of biological tasks beyond the original scope of the competition.

This first edition of the CBVCC focused on T cells exhibiting sudden changes in their migration direction, a basic behavior central to cellular migration. T cells are known to display a broad range of motility patterns, respond to chemotactic gradients, and undergo amoeboid-like migration at high speed. These characteristics make the classification task both biologically meaningful and computationally challenging. Future editions, with additional data, could expand the benchmark to include additional imaging modalities, cell types and behaviors, thereby improving the generalizability and biological relevance of the findings.

Altogether, the results highlight that the classification of cell behavior from time-lapse microscopy data remains an open field, with considerable room for improvement, especially in leveraging morphological cues and dealing with noisy spatiotemporal data. The CBVCC challenge demonstrates the importance of Open Data practices, including standardized benchmarks and the curated datasets, in bridging the gap between computer vision and cell biology, fostering collaboration across the communities, and advancing the field.

\section*{Methods}
\subsection*{Evaluation metrics and ranking criteria}
The evaluation metrics for this challenge include the area under the receiver operating characteristic curve (AUC), precision, recall, and balanced accuracy. These are defined as follows:
\begin{equation} \begin{aligned} \text{Precision} &= \frac{TP}{TP + FP}, \\ \text{Recall} &= \frac{TP}{TP + FN}, \\ \text{Balanced Accuracy} &= \frac{1}{2} \left( \frac{TP}{TP + FN} + \frac{TN}{TN + FP} \right) \end{aligned} \end{equation}
where TP, TN, FP, and FN represent the number of true positives, true negatives, false positives, and false negatives, respectively.

The final score is computed as a weighted sum of these metrics:
\begin{equation} \text{Score} = 0.4 \times \text{AUC} + 0.2 \times (\text{Precision} + \text{Recall} + \text{Balanced Accuracy}) \end{equation}
The assigned weights reflect the significance of each metric in evaluating method performance. This approach helps address the effects of class imbalances in the dataset, resulting in a fairer and more insightful assessment of the method's capabilities.

\subsection*{Cell tracking}
Cell tracks were manually generated for all the videos in the CBVCC dataset to facilitate dataset annotation and enable the evaluation of video complexity and quality metrics. Cell positions were tracked by imaging experts using the TrackMate Fiji plugin~\cite{tinevez2017trackmate, ershov2022trackmate}. Further details about the tracking process can be found in~\cite{pizzagalli2025systematic}. While the trajectories were not shared with challenge participants during the competition, they were made publicly available after its conclusion at \url{https://www.dp-lab.info/cbvcc/#dataset} for further applications of the dataset beyond the challenge.

\subsection*{Data annotation}
The data annotation process for the CBVCC challenge involved selecting regions of interest       from the original IVM videos to extract video-patches (Fig.~\ref{fig:fig2}A). Each video-patch was labeled according to the motility behavior of the cell located at the spatial and temporal center of the video-patch. This labelling design was intended to support future integration with sliding-window approaches, potentially allowing dense scanning of the original full-size IVM videos and the creation of spatially and temporally resolved maps of cellular actions across the full field of view. Class 1 video-patches were those in which the cell in the middle frame at the spatial center of the field of view exhibited a change in its direction of movement. Class 0 included cells exhibiting different behaviors or background regions.

The annotation process was performed semi-automatically, combining an initial automated pre-screening of tracks with final manual annotation. The pre-screening step classified video-patches based on two key motility metrics: net turning angle and straightness. The net turning angle was defined as the angle formed by three key points along the cell’s trajectory: the initial point of the track, the point at $t=10$ frames (corresponding to the video-patch annotation time), and the final point. Straightness~\cite{beltman2009analysing} was calculated separately for the portion of the track before the annotation ($t\leq10$ frames) and after the annotation ($t\geq10$ frames). This metric ranges from 0 to 1: a straightness of 0 indicates a non-straight trajectory, while a straightness of 1 indicates a perfectly straight trajectory. For classification purposes, an empirically fixed straightness threshold of 0.5 was applied to distinguish between straight ($>0.5$) and non-straight paths ($\leq0.5$).

Based on these metrics, video-patches were divided into two categories (Fig.~\ref{fig:fig2}B):
\begin{itemize}
    \item Class 1: video-patches representing points where the cell trajectory shows a sharp change in direction. These points were selected when the net turning angle exceeded 90 degrees, and the trajectory must show a linear movement (above the predefined straightness threshold of 0.5) both before and after the annotation point. This criterion ensures that only significant directional changes are captured, excluding minor fluctuations in the cell’s path.
    \item Class 0: this class includes three distinct categories: i) video-patches with no visible cells selected from video regions where no track centroids are present, and where only skin tissue components are visible; ii) video-patches containing cells moving along relatively straight paths, characterized by high straightness (above the predefined threshold of 0.5) both before and after the annotation point, and net turning angle less than 90 degrees; iii) video-patches where cells exhibit minimal to no movement, with low straightness (below the predefined threshold of 0.5).
\end{itemize}

After this initial automated classification, all video patches were visually inspected by a single operator and subsequently reviewed by a second operator. If a video-patch selected by this procedure did not meet the expected cellular behavior, it was either discarded or manually reassigned to the correct class based on visual evaluation. When annotation points occurred near the beginning or end of the IVM video, black frames were added to ensure that the annotation occurred in the central frame of the video-patch.

\subsection*{Video complexity and quality metrics}
To evaluate model performance as a function of the video complexity and quality, we used two metrics: the number of cells contained in the video-patches and the signal-to-noise ratio (SNR) of the video-patches. The number of cells present in each video-patch was determined by counting the number of tracks within each video-patch. The SNR was computed using a heuristic adaptation of the definition in~\cite{ulman_objective_2017}, as outlined in~\cite{pizzagalli2018leukocyte}. Let $c_{i,t}$ represent the centroid position of cell $i$ at time $t$, extracted from the tracks. For each voxel $v$ in the current frame, the distance to the nearest centroid was calculated as $d_v = \min \left( ||v - c_{i,t}|| \right)$ across all $i$. Assuming a typical cell diameter of 10 $\mu m$, each voxel $v$ was classified as foreground (FG—inside a cell) or background (BG—outside a cell) according to the following criteria:
\begin{equation}
    \begin{aligned}
        v \in \text{FG} & \quad \text{if} \quad d_v < 3 \, \mu \text{m}, \\
        v \in \text{BG} & \quad \text{if} \quad d_v > 20 \, \mu \text{m}.
    \end{aligned}
\end{equation}
The SNR was then calculated as:
\begin{equation}
SNR = \frac{|\text{avg}(FG) - \text{avg}(BG)|}{\text{std}(BG)}.
\end{equation}
where avg represents the mean pixel value, and std denotes the standard deviation of the pixel values. The SNR was computed for each frame of the video-patch, and the results were averaged to obtain a single metric per video-patch.

\subsection*{Baseline logistic regression}
Automatic segmentation of cells was performed using CellPose (v 3.1.1.1, model ``cyto3'')~\cite{stringer2025cellpose3} based on the green channel of the frames. To avoid segmentation artifacts caused by zero intensity regions, noise was added to pixels with intensity below 20. Segmentation was followed by automatic tracking using trackpy~\cite{allan2021soft}. Track post-processing addressed gaps interpolating missing frames using celltrackR~\cite{wortel2021celltrackr}. Multiple tracks within the same video-patch were handled by selecting the one closest to the video-patch midpoint, and tracks with fewer than three coordinates were discarded. Features extraction was computed through celltrackR~\cite{wortel2021celltrackr}, with features such as speed, turning angle, outreach ratio, and displacement. Features with bimodal distributions were transformed for better classification, Additional hand-crafted features were designed to specifically detect sudden changes in direction. Four logistic regression models, trained on different features sets computed from both automatic and manual tracks, were used to assess tracking performance. Model evaluation included 5-fold cross-validation to evaluate generalization ability, with performances measured by balanced accuracy and challenge final score. Additional details are provided in Supplementary Material 1.

\section*{Code and data availability}
Python code used to perform the analyses included in this paper, and the code of the participating algorithms are made available Open Source at \url{https://github.com/rcabini/CBVCC}.

Training, Validation, and Test datasets, including video patches and class-labels are publicly available at \url{https://www.dp-lab.info/cbvcc/#dataset}.

\section*{Acknowledgements}
All authors are thankful to Michael Hikey for imaging data. DUP was supported by the Swiss National Science Foundation with grant 228512 and USI FIR grant. RFC was supported by USI FIR grant and swissuniversities with RE2VITAL CHORD B grant. IMNW, DEC, KS and KBKV were supported by the AiNed Fellowship grant NGF.1607.22.020 from the Dutch Research Council (NWO). RH was funded by grants to RFG from the Natural Sciences and Engineering Research Council of Canada (418438-13), the Translational Biology and Engineering Program of the Ted Rogers Centre for Heart Research, and the Canadian Institutes of Health Research (156279 and 186188). RFG is the Canada Research Chair in Quantitative Cell Biology and Morphogenesis.

\bibliographystyle{IEEEtran} 
\bibliography{bibliography}
\end{document}


\section*{Supplementary Material 1 --\\Baseline logistic regression model}

\paragraph{Tracking.}
Automatic segmentation was performed using Cellpose~\cite{stringer2025cellpose3} (model ``cyto3'') based on the green channel as follows. Because some images contained regions of zero intensity that were causing artifacts during segmentation, we added noise to pixels with intensity $I_0(p)$ below 20 (in range 0--255), as:
\[
I(p)=
\begin{cases}
I_0(p), & I_0(p) > 20, \\
q_{20} + \varepsilon(p), & I_0(p) \le 20,
\end{cases}
\]
with $q_{20}$ the 20\% quantile of unique $I_0$ values, and $\varepsilon(p) \sim \mathcal{N}(0,5)$. Segmentation was performed using Cellpose (v3.1.1.1, flow\_threshold $=0.4$, cellprob\_threshold $=2$, max\_size\_fraction $=0.08$, using default normalization) in Python (v3.12.2). Centroids were generated from segmentation masks using scikit-image~\cite{van2014scikit} (v0.25.0), and tracks were automatically generated using trackpy (v0.6.4, search range $=20$, memory $=30$—essentially allowing all links within the limited-size patch). We did not extensively optimize these settings as the goal was to evaluate how off-the-shelf tracking performs in this context. In addition to automated tracking results, we also used manually generated tracks for each video-patch.

\paragraph{Track processing.}
Automatically or manually generated tracks can contain gaps (intermediate frames for which no cell is detected). Before feature extraction, these gaps were detected and positions interpolated using celltrackR~\cite{wortel2021celltrackr} (v1.2.0, R v4.4.1, method interpolateTrack with how = ``spline'', \url{https://ingewortel.github.io/celltrackR}). For some video-patches, manual or automated tracking resulted in multiple tracks as there were multiple cells in the video-patch. In those cases, we selected the track that was closest to the video-patch midpoint $(25,25)$ in the middle frame of the image sequence. Tracks with less than 3 coordinates were discarded, since a “change in behavior” is undefined when there is only one movement step (2 coordinates).

\paragraph{Feature extraction.}
The following track features were extracted from each of the selected processed tracks using default methods from celltrackR~\cite{wortel2021celltrackr} (v1.2.0, R v4.4.1): speed, meanTurningAngle, outreachRatio, displacementRatio, straightness, asphericity, and displacement (see package documentation for details). Because exploratory data analysis on the training data (Fig.~\ref{fig:figS1}) showed that the features straightness and outreachRatio were bimodal in class 0, with class 1 unimodal somewhere in between, we transformed these feature values as: $v_{\text{new}} = \lvert v_{\text{orig}} - b \rvert$, with $b=0.55$ for outreachRatio and $b=0.45$ for straightness  based on manual inspection of the distributions (Fig.~\ref{fig:figS1}). In addition, we included two additional standard features: $N_{\text{coordinates}}$ in the track (as tracks of very few coordinates may indicate that there was no real cell, therefore pointing to class 0), and the track’s distance $d_{\text{center}}$ at frame 10 (where again a large distance/absence of a cell in that frame may indicate that the video-patch was not focused on a cell and therefore class 0).

In addition to the abovementioned “basic” features, which used little information about the specific task, we designed two hand-crafted features to detect (sharp) changes in direction based on the step displacement vectors $\vec{v}_t = \vec{x}_{t+1} - \vec{x}_t$  (where $\vec{x}_t$ is the cell’s position at time t). The first feature, $\Delta\theta$, uses the angle $\theta_t$ (w.r.t. the positive x-axis) of each step $\vec{v}_t$ , taking the difference in average direction before and after the middle of the video $t_{mid}$ (frame 10):
\[
\Delta\theta =
\left|
\langle \theta_t \rangle_{t \ge t_{\text{mid}}}
-
\langle \theta_t \rangle_{t < t_{\text{mid}}}
\right|,
\]
The second feature instead uses the normalized average displacement vectors before and after $t_{mid}$:
\[
s_{\text{norm}} =
\left\|
\frac{\vec{v}_{t < t_{\text{mid}}}}{\lVert \vec{v}_{t < t_{\text{mid}}} \rVert}
+
\frac{\vec{v}_{t \ge t_{\text{mid}}}}{\lVert \vec{v}_{t \ge t_{\text{mid}}} \rVert}
\right\|,
\]
with
\[
\vec{v}_{t < t_{\text{mid}}}
=
\frac{1}{N_{t < t_{\text{mid}}}}
\sum_{t=t_{\min}}^{t_{\text{mid}}}
\vec{v}_t,
\quad
\vec{v}_{t \ge t_{\text{mid}}}
=
\frac{1}{N_{t \ge t_{\text{mid}}}}
\sum_{t=t_{\text{mid}}}^{t_{\max}-1}
\vec{v}_t.
\]
where $t_{min}$ and $t_{max}$ are the first and last time point of the track, respectively. The value of $S_{norm}$ is small when the two vectors point in opposite directions (cancelling each other out), and attains its max value of 2 for straight, ballistic motion. Note that the features $d_{center}$, $\Delta\theta$, and $S_{norm}$ are only defined when there is a cell observed at $t = t_{mid}$. When this is not the case, we consider this as evidence that there is no focus cell in the video-patch, and therefore set these features to “class 0-like” values (Fig.~\ref{fig:figS1}): $d_{center} = 50$, $\Delta\theta = 0$, $S_{norm} = 2$. 

\paragraph{Logistic regression.}
To compare the performance of logistic regression using automated versus manual tracks and using all features versus only those not specifically crafted for the task, we trained four logistic regression models (Table~\ref{tab:logreg}) using scikit-learn (v1.6.1) in python (v3.12.2), using L2 regularization (C=200). In cases where all features were missing because no track was observed, all features were set to $-1$ before training and prediction (and an extra binary feature was added to indicate track missingness, which was used for all models in Table~\ref{tab:logreg}).

\begin{table}[h]
\centering
\caption{Overview of logistic regression models.}
\label{tab:logreg}
\begin{tabular}{lll}
\hline
Model & Tracks & Trained on \\
\hline
auto-all & Automated & All features \\
auto-basic & Automated & All features except $\Delta\theta$ and $S_{\text{norm}}$ \\
manual-all & Manual & All features \\
manual-basic & Manual & All features except $\Delta\theta$ and $S_{\text{norm}}$ \\
\hline
\end{tabular}
\end{table}

\paragraph{Model evaluation.}
All models were evaluated on the held-out test data (validation and test datasets). To further assess generalization ability, we performed 5-fold cross-validation on the training data (ensuring that video-patches from the same original video only occurred in the training or test fold, but not both). Because of the large variation observed in test accuracy between folds, we repeated this process 5 times with different splits to obtain a better performance estimate. All performances are reported as balanced accuracy (from scikit-learn) and overall score.

\clearpage
\section*{Supplementary Material 2 -- \\Description of the participating algorithms}

\subsection*{GIMR -- Garvan Institute of Medical Research}
The GIMR team developed TrajNet, a track-based approach designed for analyzing and classifying CBVCC data. It integrates multiple components, including cell tracking, feature extraction using a convolutional neural network (CNN), an attention-based track selection module, and a multilayer perceptron (MLP) classifier.

The first step involves tracking cells across video frames. A retrained Cellpose model (based on cytotorch0) is employed for segmentation~\cite{stringer2025cellpose3}, combined with trackpy/laptrack for track generation~\cite{allan2021soft}. To optimize performance and minimize noise, only the green channel of each video frame is processed. This results in multiple trajectories per video, each representing a cell’s movement over time. To mitigate the effect of segmentation and tracking errors, only high-intensity ($>50$) and large ($>50$ pixels) cells are considered.

Trajectories are then zero-padded to a fixed 3 $\times$ 20 size, representing (x, y, t) coordinates over 20 frames. A shallow two-layer CNN is employed for feature extraction to capture motion characteristics. The network consists of two 1D convolutional layers with ReLU activation and batch normalization, followed by max-pooling and average-pooling operations. This simple architecture is motivated by the insight that motion features, such as speed and turning angles, can be effectively captured through first- and second-order differential operations.

To focus on the most informative trajectories in the video-patch, an attention mechanism assigns weights to each trajectory’s features allowing the model to aggregate the most representative motion patterns through weighted averaging. This enables the model to prioritize salient motion patterns, akin to human visual assessment. The aggregated features are then fed into an MLP classifier, which is trained jointly with the attention module for end-to-end classification.

Overfitting is mitigated through reduced model complexity, dropout layers with a rate of 0.5, weight decay, and data augmentation techniques such as rotation, scaling, time reversal, speed adjustments, and interpolation. Final predictions are obtained via model ensembling by averaging the outputs of the top three performing models.
The source code for TrajNet is available at \url{https://github.com/lxfhfut/TrajNet}.

\subsection*{LRI Imaging Core -- Cleveland Clinic}
The LRI team developed the CB pipeline, a track-based approach based on an extension of the LabGym tool~\cite{hu2023labgym, goss2024quantifying}, designed specifically for analyzing and classifying CBVCC data. Since multiple cells can be present in a video-patch, each exhibiting different behaviors over time, the CB pipeline ensures that all cells within the video-patches are tracked, and their behaviors are classified at every frame. Specifically, cell behaviors have been categorized into three predefined motion patterns: inplace (minimal movement), linear (consistent movement in one direction), and orient (sudden directional change). The pipeline integrates three main components: cell tracking, feature extraction using CNNs and long-short-term-memory (LSTM) layers~\cite{hochreiter1997long}, and an MLP classifier.

Cell tracking is performed using LabGym’s Detector module, which is based on Detectron2~\cite{wu2019detectron2}. To train the Detector, 475 manually segmented frames were augmented to produce 9850 training images. The tracking method assigns a unique identity to each cell and links them across frames using Euclidean distance calculations, ensuring reliable trajectory continuity.

For behavior classification, a model was trained on the three predefined motion patterns using manually annotated data for supervised learning. The classification pipeline consists of three components: an Animation Analyzer, a Pattern Recognizer, and a Decision Maker. The Animation Analyzer processes video-patches, cropped around each cell using segmentation masks, with 2D convolutional and LSTM layers~ \cite{hochreiter1997long}. The Pattern Recognizer extracts spatial features using 2D convolutional layers from a 2D image projection of each cell’s boundary across all frames. The Decision Maker then combines these outputs through an MLP with a SoftMax function to predict behavior probabilities.

For the CBVCC classification task, the objective is to distinguish video-patches into two classes. A locational filter is applied to focus only on cells passing through the center of the frame. Predictions from frames 14, 15, and 16 are used, with weights assigned as follows: 0.1429 for linear behavior, 0.2857 for inplace behavior, and 0.5714 for orient behavior. The final summary prediction score determines whether a video belongs to class 0 or class 1.

To mitigate overfitting, the model incorporates dropout layers with a 0.5 rate, and extensive data augmentation techniques, including rotation, flipping, and brightness adjustments. Training optimization follows a categorical cross-entropy loss function with an adaptive learning rate schedule.

The source code for the CB pipeline is available at \url{https://github.com/yujiahu415/CBVCC}.

\subsection*{QuantMorph -- University of Toronto}
The QuantMorph team developed a track-based approach for analyzing and classifying CBVCC data. This method uses the pre-trained Segment Anything Model 2 (SAM2)~\cite{ravi2024sam} for cell segmentation and tracking, followed by a custom LSTM-based neural network for classifying the cell tracks~\cite{hochreiter1997long}.

For data preprocessing, only the green channel of each frame was used. Cell segmentation and tracking were performed using the pretrained SAM2, which segments and tracks objects in a video starting from the centroid in the first frame they appear. Centroids were initially identified in each frame using a Laplacian of Gradients blob detector~\cite{van2014scikit}. The segmentation process was terminated if SAM2 failed to produce a segmentation within 10 pixels of the previous frame’s centroid or if the segmented area exceeded 500 pixels.

In the training set, manual pruning was applied to class 1 videos to ensure only class 1 tracks were used for training. Class 0 videos, containing only class 0 objects, retained all tracks for training. The centroid coordinates across frames were used to construct a 2 $\times$ 20 matrix representing the ($x, y$) coordinates of each object’s trajectory. If any object was not segmented across all frames, missing data was filled by assigning the centroid from either the first or last segmented frame.

The classifier architecture consists of a 1D convolutional layer followed by five LSTM layers~\cite{hochreiter1997long}, and a final fully connected layer with SoftMax activation for classification.

During inference, a video-patch was classified as class 1 if any segmented object track within the video-patch was classified as class 1. Videos with no predicted class 1 object tracks, or with no successfully segmented objects, were classified as class 0. The object with the highest class 1 probability was used as the probability for classifying the entire video-patch.

To mitigate overfitting, the model includes dropout layers with a rate of 0.7, along with data augmentation by training on both forward and reverse orientations of centroid tracks. The training optimization uses a categorical cross-entropy loss function with class weights to address class imbalance, along with an adaptive learning rate schedule.

The source code is available at \url{https://bitbucket.org/raymond_hawkins_utor/cbvcc/src/main/}.

\subsection*{dp-lab -- USI}
The dp-lab team developed an end-to-end deep learning (DL) approach for analyzing and classifying CBVCC data. This approach is based on a 3D CNN for video classification tasks, which simultaneously processes spatial and temporal components~\cite{ji20123d}. No explicit cell segmentation or tracking was performed.

The classification model consists of three convolutional blocks, each containing two 3D convolutional layers with a progressively decreasing number of filters. The first block has 128 filters, the second has 64, and the third has 32. Each convolutional layer is followed by batch normalization and ReLU activation. After each block, a max pooling layer is applied to downsample the feature maps. The output from the final max pooling layer is flattened and passed through a MLP with two fully connected layers, each followed by ReLU activation and dropout layers. The final classification is performed using a fully connected output layer with a SoftMax activation function, producing a probability distribution over the two classes.

To mitigate overfitting, the model incorporates reduced complexity, dropout layers with a 0.2 rate, and data augmentation techniques, including random rotation, zooming, vertical and horizontal flipping, shifting, and contrast stretching. The training optimization uses a categorical cross-entropy loss function with class weights to address class imbalance.

The source code is available at \url{https://github.com/rcabini/CBVCC_CNN}.

\subsection*{UWT-SET -- University of Washington}
The University of Washington (UWT-SET) team developed the Enhanced Swin-Tiny Model, an end-to-end DL framework specifically designed for analyzing and classifying CBVCC data. This approach integrates two primary components: a standard Video Shifted Window Transformer Tiny (Swin-Tiny)~\cite{liu2021swin} and a motion-guided data augmentation pipeline.

The input video frames are resized to a 224 $\times$ 224 resolution. The classification model, Swin-Tiny, is a parameter-efficient variant of the original Swin Transformer, designed to balance computational efficiency with the capacity to capture subtle motion. Its hierarchical architecture consists of four stages with layer depths of [2, 2, 6, 2], an embedding dimension of 96, and multi-head self-attention layers featuring 3, 6, 12, and 24 heads across successive stages. The model employs a patch size of [2, 4, 4] and a drop path rate of 0.2, with layer normalization and GELU activations applied to each block.

To mitigate overfitting, training is augmented with a motion-guided data augmentation pipeline designed to enhance both cell movement and imaging conditions. Frame differencing identifies high-motion regions, while HSV-based filtering selects cells from these regions by discarding irrelevant pixels based on color and brightness. Local affine transformations are applied to these high-motion patches to simulate variations in cell trajectories. Additionally, the augmentation pipeline applies global transformations across the entire frame, such as color jitter and Gaussian blur, to replicate variations in imaging conditions.

The model is trained using the AdamW~\cite{loshchilov2017decoupled} optimizer with a weight decay of 0.02 and a cosine annealing learning rate schedule. To address class imbalance, class weighting is incorporated into the cross-entropy loss function.

\subsection*{Computational Immunology -- Radboud University}

The Computational Immunology team developed an ensemble of a track-based method and an end-to-end DL image-based method. The final prediction was obtained by averaging the outputs of both models. The track-based method was built on a logistic regression model, as described in Supplementary Materials 1, trained on all features extracted from automatically generated tracks. The end-to-end DL image-based method is based on~\cite{schneider2023learnable} and employs a 3D convolutional encoder with contrastive learning to learn a low-dimensional feature representation, which is then fed to an MLP classifier, as described below.

Contrastive learning compares reference training input videos with 1 positive anchor ($y^+$) and multiple negative anchors ($y_{1\ldots n}^-$) to learn an embedding space in which similar samples $(x,y^+)$ are pulled towards each other and dissimilar ones $(x,y_i^-)$ are pushed apart. Following~\cite{schneider2023learnable}, training data $X$ was first ordered by class. During training, a batch of $(b = 8)$ reference videos are sampled and mapped to the 2-dimensional latent space ($f(x)$, dimension $8 \times 2$). The corresponding $b = 8$ positive anchors are sampled with an offset of one: $\forall x_k \in X \rightarrow y^+ = x_{k+1}$. Negative anchors $y_{1\ldots b}^-$ are sampled uniformly from $X \setminus \{x, y^+\}$. A Siamese encoder network $f(x)$ and $f'(y)$ is optimized to map reference $x$ and anchors $y$ to the shared low dimensional space. The encoder architecture consists of two 3D convolutional layers followed by max-pooling layers, which are then passed through two fully connected layers. Using cosine similarity $\psi$ and writing $\psi\left(f(x), f'(y)\right)$ as $\psi(x, y)$ for simplicity, the following contrastive loss~\cite{schneider2023learnable}, 2023) is minimized:
\[
L_{\text{contrastive}} = \mathbb{E}_{x,y^+,y^-_{1,\ldots,n}} \left[ - \psi(x,y^+) + \log \sum_{i=1}^{b} e^{\psi(x,y_i^-)} \right] 
\]
The learned low-dimensional features are then fed to an MLP for classification, which consists of an elementwise activation layer with Tanh, followed by a fully connected Softmax layer for the final classification output. The entire network is optimized jointly on the representation ($L_{\text{contrastive}}$) and classification ($L_{\text{BCE}}$, cross entropy) objectives with equal weights: $L_{\text{contrastive}} + L_{\text{BCE}}$, using Adam optimization~\cite{kingma2014adam} (250 epochs, learning rate $3 \times 10^{-4}$). No data augmentation was performed for this method.

The source code is available at \url{https://github.com/DavidCicch/CBVC_Challenge}.

\subsection*{BioVision -- University of Central Florida}
The BioVision team developed an end-to-end DL approach using the Video Shifted Window Transformer (Swin) model for analyzing and classifying CBVCC data~\cite{liu2021swin}. The model was pretrained on the Something-Something v2 (STHv2) dataset, which focuses on human-object interactions from an egocentric perspective~\cite{mahdisoltani2018effectiveness}. This pre-training helps the model capture fine-grained, localized actions, making it well-suited for analyzing dynamic cell movements and filament interactions in microscopy videos. Further fine-tuning on the CBVCC training dataset allowed the model to adapt to domain-specific features.

Input video frames were resized to 224 $\times$ 224 resolution and normalized using the standard preprocessing values of the STHv2 dataset. No explicit cell segmentation or tracking was performed.

The classification model is based on a Swin backbone with a hierarchical architecture consisting of four stages with depths of [2, 2, 18, 2]. It uses an embedding dimension of 128 and multi-head self-attention with 4, 8, 16, and 32 heads across stages. The model employs a patch size of [2, 4, 4], a window size of [16, 7, 7], a drop path rate of 0.2, and patch normalization for improved learning efficiency. After feature extraction, 3D adaptive average pooling is applied to reduce dimensionality while preserving key spatial and temporal features. The final classification is performed using a fully connected output layer with a softmax activation function.

Training optimization follows a categorical cross-entropy loss function with an adaptive learning rate schedule.

The source code is available at \url{https://github.com/jkini/CellBehaviorVideoClassification}. 

\section*{Supplementary figures}

{\centering\includegraphics[width=\textwidth]{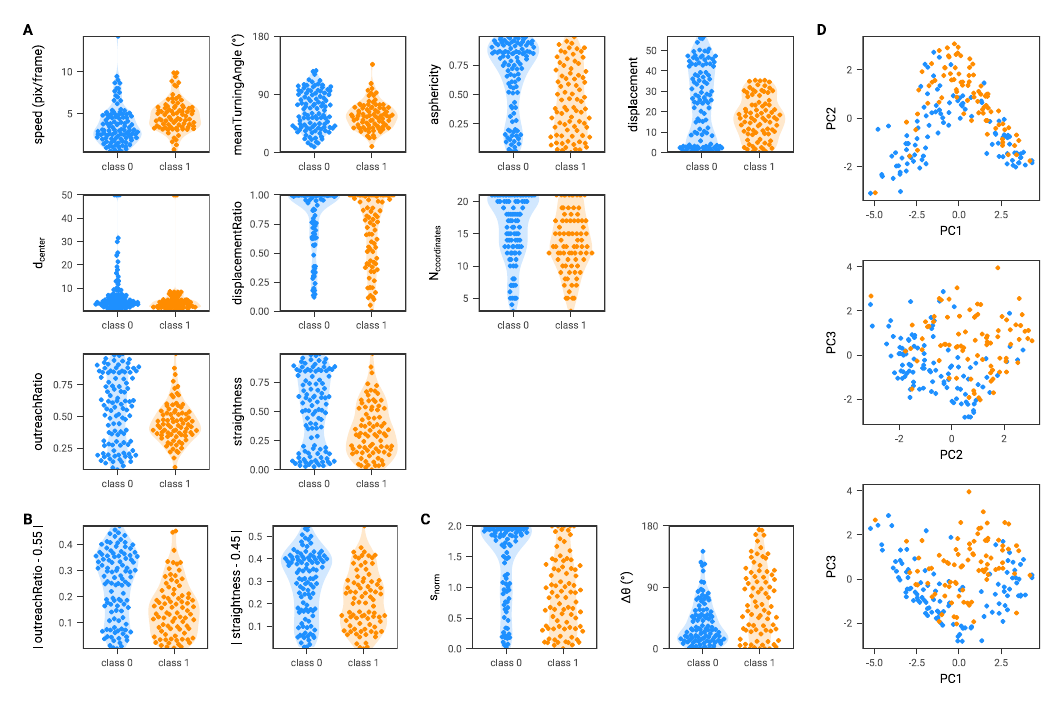}
\captionof{figure}{ \footnotesize \textbf{Exploratory data analysis on manual track features (training data).} \textbf{A.} “Basic” features. \textbf{B.} Transformed versions of outreachRatio and straightness used for logistic regression. \textbf{C.} Hand-crafted features. \textbf{D.} PCA projections along the first three principal components using all features.}\label{fig:figS1}}

\clearpage
{\centering\includegraphics[width=\textwidth]{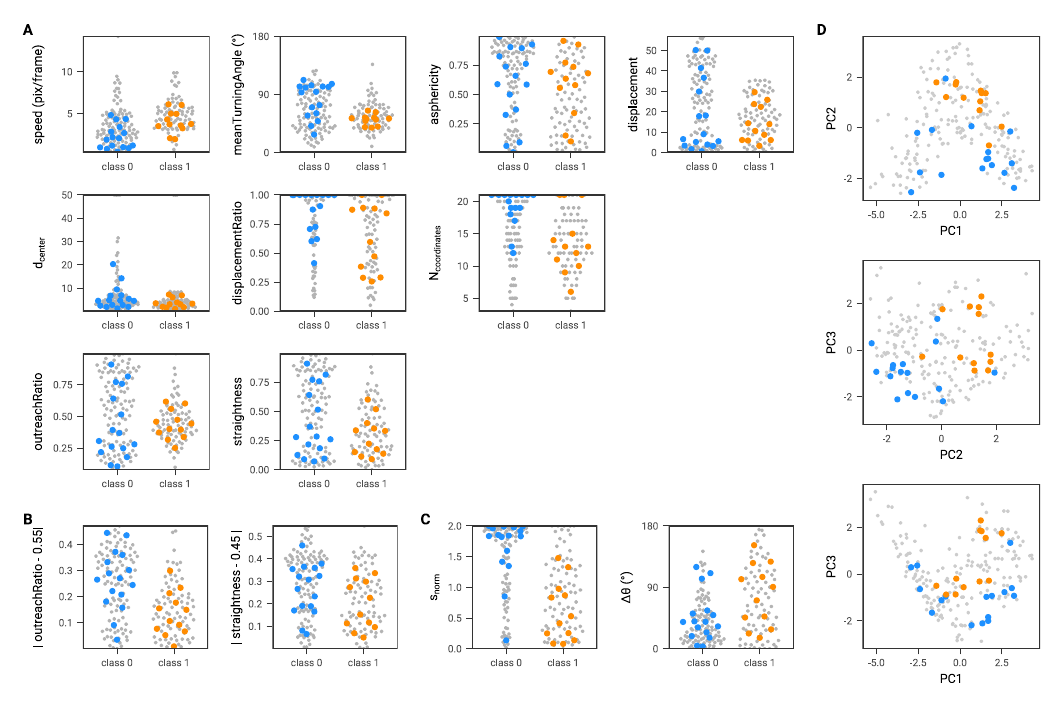}
\captionof{figure}{ \footnotesize \textbf{Exploratory data analysis on manual track features (validation set).} Gray dots: training data from Figure S1. \textbf{A.} “Basic” features. \textbf{B.} transformed versions of outreachRatio and straightness used for logistic regression. \textbf{C.} Hand-crafted features. \textbf{D.} PCA projections along the first three principal components using all features.}\label{fig:figS2}}

{\centering\includegraphics[width=0.9\textwidth]{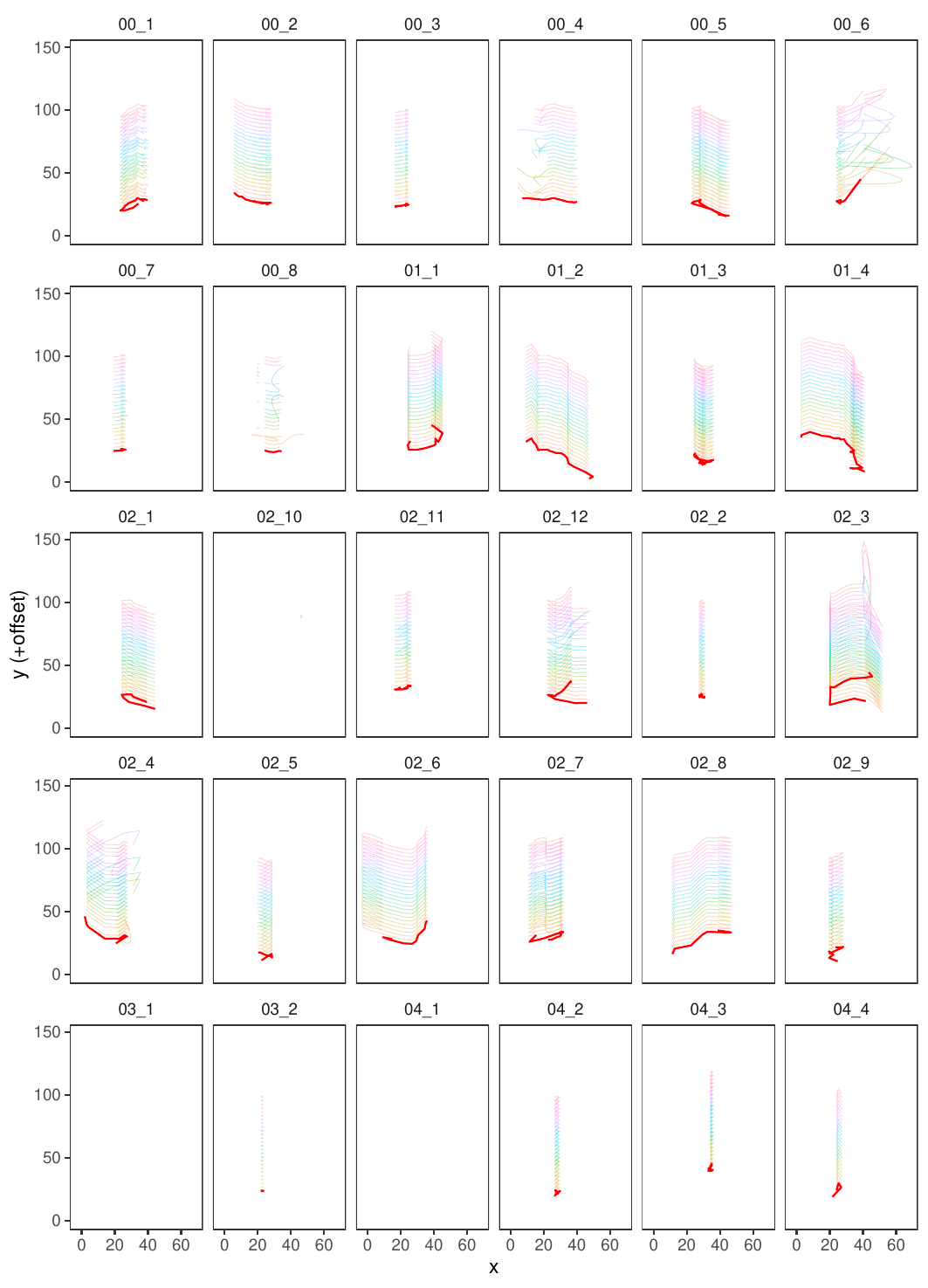}
\captionof{figure}{ \footnotesize \textbf{Variation in automated tracking results on the validation set.} Thick, red line: tracks used for evaluation. Thin colored lines: results from 25 replicates of the tracking pipeline with different random seeds.}\label{fig:figS3}}

{\centering\includegraphics[width=\textwidth]{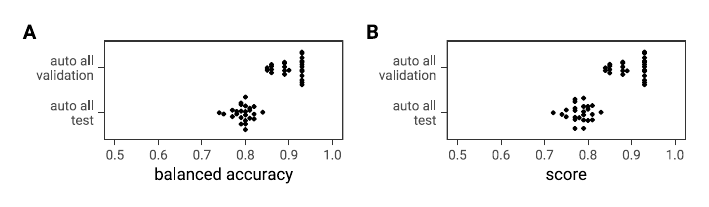}
\captionof{figure}{ \footnotesize \textbf{Effect of errors in automated tracking on logistic regression performance.} The automated tracking pipeline was repeated 25 times to generate different tracks for validation and test sets (Fig.~\ref{fig:figS3}), used to evaluate a single trained model (“auto all” from Figure 6, trained on the full training set). The figure shows evaluation results across those 25 tracking replicates (dots). Differences in generated tracks translated to substantial variation in model performance – even though both the trained model and the videos used for validation were fixed.}\label{fig:figS4}}

\bibliographystyle{IEEEtran} 
\bibliography{bibliography}